\documentclass[prl,twocolumn,superscriptaddress]{revtex4-1}
\usepackage{amssymb}
\usepackage{amsfonts}
\usepackage{mathrsfs}
\usepackage{graphics,graphicx,epsfig,amsmath,amsthm}
\usepackage{floatrow}
\usepackage{caption}
\usepackage[labelformat=simple]{subfig}

\usepackage{bm}
\usepackage{bbm}
\usepackage{multirow}
\usepackage{array}
\usepackage{color}
\usepackage{cases}
\usepackage{booktabs }
\usepackage[usenames,dvipsnames]{xcolor}

\usepackage{xcolor}
 
\usepackage{float}
\usepackage[a4paper,colorlinks=true,
linkcolor=blue,citecolor=blue,
pdfauthor={ },
pdftitle={ },
pdfsubject={ },
pdfkeywords={ }]{hyperref}

\begin{document}

\title{Experimental violation of the Leggett-Garg inequality with a single-spin system}
 
\author{Maimaitiyiming Tusun}
\affiliation{CAS Key Laboratory of Microscale Magnetic Resonance and School of Physical Sciences, University of Science and Technology of China, Hefei 230026, China}
\affiliation{CAS Center for Excellence in Quantum Information and Quantum Physics, University of Science and Technology of China, Hefei 230026, China}
\affiliation{School of Physics and Electronic Engineering, Xinjiang Normal University, Urumqi 830054, China}

\author{Wei Cheng}
\affiliation{CAS Key Laboratory of Microscale Magnetic Resonance and School of Physical Sciences, University of Science and Technology of China, Hefei 230026, China}
\affiliation{CAS Center for Excellence in Quantum Information and Quantum Physics, University of Science and Technology of China, Hefei 230026, China}
 
\author{Zihua Chai}
\affiliation{CAS Key Laboratory of Microscale Magnetic Resonance and School of Physical Sciences, University of Science and Technology of China, Hefei 230026, China}
\affiliation{CAS Center for Excellence in Quantum Information and Quantum Physics, University of Science and Technology of China, Hefei 230026, China}
 
\author{Yang Wu}
\affiliation{CAS Key Laboratory of Microscale Magnetic Resonance and School of Physical Sciences, University of Science and Technology of China, Hefei 230026, China}
\affiliation{CAS Center for Excellence in Quantum Information and Quantum Physics, University of Science and Technology of China, Hefei 230026, China}
 
\author{Ya Wang}
\affiliation{CAS Key Laboratory of Microscale Magnetic Resonance and School of Physical Sciences, University of Science and Technology of China, Hefei 230026, China}
\affiliation{CAS Center for Excellence in Quantum Information and Quantum Physics, University of Science and Technology of China, Hefei 230026, China}
 
\author{Xing Rong}
\email{xrong@ustc.edu.cn}
\affiliation{CAS Key Laboratory of Microscale Magnetic Resonance and School of Physical Sciences, University of Science and Technology of China, Hefei 230026, China}
\affiliation{CAS Center for Excellence in Quantum Information and Quantum Physics, University of Science and Technology of China, Hefei 230026, China}
 
\author{Jiangfeng Du}
\email{djf@ustc.edu.cn}
\affiliation{CAS Key Laboratory of Microscale Magnetic Resonance and School of Physical Sciences, University of Science and Technology of China, Hefei 230026, China}
\affiliation{CAS Center for Excellence in Quantum Information and Quantum Physics, University of Science and Technology of China, Hefei 230026, China}
 
\begin{abstract}
Investigation the boundary between quantum mechanical description and classical realistic view is of fundamental importance.
The Leggett-Garg inequality provides a criterion to distinguish between quantum systems and classical systems, and can be used to prove the macroscopic superposition state.
A larger upper bound of the LG function can be obtained in a multi-level system.
Here, we present an experimental violation of the Leggett-Garg inequality in a three-level system using nitrogen-vacancy center in diamond by ideal negative result measurement.
The experimental maximum value of Leggett-Garg function is $K_{3}^{exp}=1.625\pm0.022$ which exceeds the L$\mathrm{\ddot{u}}$ders bound with a $5\sigma$ level of confidence.
\end{abstract}
 
\maketitle
 
The objects we can see with our naked eye are composed of a very large number of atoms, and these atoms can be in a quantum superposition state. If the laws of quantum mechanics are extrapolating to the laws satisfied by everyday objects, there is inevitably the prospect of macroscopic coherence. Schr$\mathrm{\ddot{o}}$dinger's cat, simultaneously both dead and alive, is considered to be one of the most characteristic examples of  macroscopic coherence. This situation runs counter to our intuitive understanding of how the daily macroworld works.

Leggett and Garg gave a method to verify whether a given system can be in a superposition of discrete states\cite{Leggett_1985}.
They proposed two assumptions that the classical world certainly satisfied\cite{Leggett_2002,Leggett_2008}: (i) \emph{macroscopic realism} (MR)--- that a system cannot be in a superposition of the classically observable state and the measurement of the object will get a certain value; (ii) \emph{Non-invasive measurability} (NIM)---that it is possible in principle to determine the state of a system with arbitrarily precision without destroying its subsequent evolution. Classical world meet the requirements of these assumptions, but for quantum mechanics, neither of these two assumptions can be satisfied. The Leggett-Garg inequality (LGI) is a test of MR under a set of reasonable assumptions about the system, in particular a version of NIM.
The violation of the inequality leads to the conclusion that either MR or the other assumptions is wrong\cite{Maroney_2014}.
In this way, the LGI provide a method to investigate the existence of macroscopic coherence and to test the applicability of quantum mechanics as we scale from microscopic to macroscopic world\cite{Leggett_2002}.

Until recently it has been believed that the maximal violation of three-time measurement LGI, which characterized the correlation between the results of this measurements, is independent of the number of possible macroscopically distinct states of the system. This is due to the fact that the measurements are dichotomic\cite{Katiyar_2017}.
However, Budroni and Emary\cite{Budroni_2014} showed that this is only true if the measurement follows the naive L$\mathrm{\ddot{u}}$ders update rule\cite{luder_1951}. According to L$\mathrm{\ddot{u}}$ders rule, the state is updated as $\rho\rightarrow\Pi_{\pm}\rho\Pi_{\pm}$  depending on the outcome of the measurement, where $\Pi_{+}$ and $\Pi_{-}$ are two projectors on eigenspaces associated with measurement results $Q_+=1$ and $Q_-=-1$.
In a more general case, one can observe greater violations by using a higher-dimensional system.
Therefore, striving for a larger violation value in a higher-dimensional system is worth studying.
The quantum upper bound of the LGI depends on both the details of the measurement and the dimension of the system\cite{Katiyar_2017}.
For the LGI experiment performed on the three-level system, a higher theoretical quantum upper bound can be obtained than that on the two-level system\cite{Palacios-Laloy_2010,Katiyar_2013,Dressel_2011}.
With interest in the beyond quantum upper bound of the LGI, the NMR system\cite{Katiyar_2017} and single photon\cite{Wang_2017} are used to mimic the three-level system.
There is also a work performed in a single three-level system\cite{George_2013}. However, this work used the projective measurement which is invasive measurement method and fails to convince a macrorealist.
Achieving the violation of the LGI which breaks through the L$\mathrm{\ddot{u}}$ders bound in a single three-level system remains elusive.
Here, we report an experimental violation of LGI with the exceeding of L$\mathrm{\ddot{u}}$ders bound in a real three-level system.
Our experiment is performed in a single nitrogen-vacancy (NV) center in diamond.
The Ideal Negative Result Measurement (INRM) method is implemented to avoid invasive measurements\cite{Katiyar_2017,Knee_2012}.
 
\begin{figure}
\centering
\includegraphics[width=.9\textwidth]{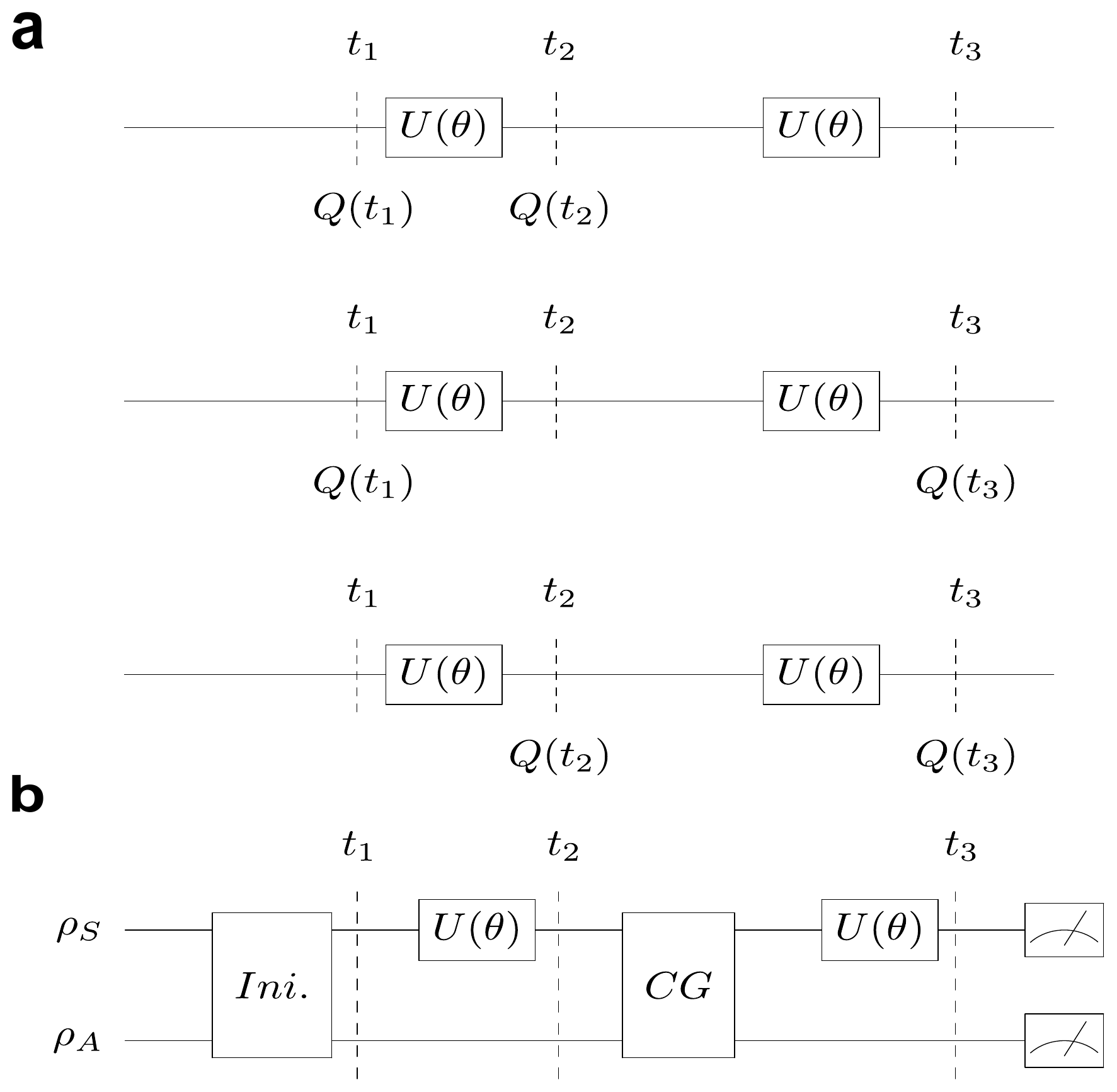}
\caption{(\textbf{a}) Three-time measurement of LG test, in this scenario, the initialization can be regard as the first measurement. (\textbf{b}) General scheme for the LG test with INRM. The state of the system is initialized to $|\psi(t_{1})\rangle$, and the ancilla qubit is initialized to $|0\rangle$. After first and second evolution of $U(\theta)$, the state of the system is evolved to $|\psi(t_{2})\rangle$ and $|\psi(t_{3})\rangle$, corresponding to observable dichotomic variables $Q(t_2)$ and $Q(t_3)$. INRM is implemented by controlled gate (CG) and the postselecting the final state populations corresponding to the non-flipping of the ancilla qubit.}
\label{1}
\end{figure}

The LGI test the correlations of a single system measured at different times. The most frequently encountered inequalities concern the n-measurement LG strings\cite{Budroni_2014}
\begin{equation}
\begin{split}
  K_{n}=\langle Q(t_{2})Q(t_{1})\rangle + \langle Q(t_{3})Q(t_{2})\rangle+ \langle Q(t_{4})Q(t_{3})\rangle\\
  +\cdot\cdot\cdot+\langle Q(t_{n})Q(t_{n-1})\rangle-\langle Q(t_{n})Q(t_{1})\rangle.
\end{split}
\end{equation}
Here, $Q$ is a macroscopic dichotomic variable with distinct outcomes $q$, at some time $t_i$, the outcomes of these measurements are denoted as $q_{i}^{(1)}=+1$ and $q_{i}^{(2)}=-1$, $n$ denotes the number of time steps. In a macrorealistic system, the outcomes $q_{i}^{(l)}(l=1,2)$ represent the two distinct sates of the system. The correlation function is $\langle Q_{i}Q_{j}\rangle=\sum_{l,m}q_{i}^{(l)}q_{j}^{(m)}P(q_{i}^{(l)},q_{j}^{(m)})$, where $P(q_{i}^{(l)},q_{j}^{(m)})$ is the joint probability to obtain $q_{i}$ and $q_{j}$ at times $t_{i}$ and $t_{j}$. The n-measurement LGIs are bounded as follows. If $n\geq 3$ and odd then $-n\leq K_{n}\leq n-2$, and if $n\geq 4$ and even then $-(n-2)\leq K_{n}\leq n-2$\cite{Budroni_2014}.
In the typical three-time measurement LGI scenario as shown in Fig.~\ref{1}(\textbf{a}), the LG function is given by\cite{Emary_2014}
\begin{equation}
  K_{3}=\langle Q(t_{2})Q(t_{1})\rangle + \langle Q(t_{3})Q(t_{2})\rangle-\langle Q(t_{3})Q(t_{1})\rangle.
\end{equation}
More specifically, when the measurements performed to the system do not disturb the system (NIM assumption) and the system is classical (macrorealistic), the value of $K_{3}$ is bounded by $-3\leq K_{3}\leq1$. On the other hand, when the system is quantum, it is possible to choose the evolution time between measurements in such a way that $K_{3}$ will go beyond 1, violating the LGI that $K_{3}\leq 1$\cite{Emary_2014}. Quantum mechanical experiments are able to violate three-time measurement LGI up to 1.5 for a qubit with dichotomic measurements $q_{i}^{(1,2)}=\pm1$. More general systems have the same bound when the measurements follow the L$\mathrm{\ddot{u}}$ders update rule. However, if one relaxes the assumption that the measurement follows the L$\mathrm{\ddot{u}}$ders update rule, the quantum bound on $K_{3}$ could be extended to a value that depends on the dimension of the system\cite{Budroni_2014}.
For a three-level system, the measurement operators are $\Pi_1=|1\rangle_{n}$\!$\langle1|$, $\Pi_0=|0\rangle_{n}$\!$\langle0|$ and $\Pi_{-1}=|\textnormal{-}1\rangle_{n}$\!$\langle\textnormal{-}1|$ corresponding to the three states of system. $\Pi_1$ and $\Pi_0$ are associated with outcome $q^{(1)}=q^{(0)}=1$. $\Pi_{-1}$ is associated with outcome $q^{(-1)}=-1$.
The quantum state after the measurement is updated as $\rho \rightarrow\Pi_1\rho\Pi_1+\Pi_0\rho\Pi_0$ or $\Pi_{-1}\rho\Pi_{-1}$ depending on whether the measurement result is $+1$ or $-1$. This protocol is different from the L$\rm {\ddot{u}}$ders rule and can lead to the value $K_{3}=1.756$ in the ideal case.
 
Ideal negative result measurement is one of candidates to achieve non-invasive measurement. Measurement process can be regarded as an interaction process between the system to be measured and the apparatus. For instance, a detector couples with one of system states, registering a ‘click’ if the system is in the specific state, but it does not interact with the other states. If the detector clicks, the result is discarded, and if it does not click, we keep the result. In this way, we can infer system states without any interaction between the system to be measured and the apparatus by keeping only the negative (no click) results.
In our scheme, the INRM scheme is realized by controlled gates and the postselection of the ancilla qubit state as depicted in Fig.~\ref{1}(\textbf{b}).
The ancillary qubit is initialized in the state $|0\rangle$.
The controlled gates flip the ancilla qubit except when the three-level system is in one of the three states.
Postselecting the situation that the ancilla qubit has not been flipped, the non-invasiveness of the measurement can be guaranteed.
 
\begin{figure}
\centering
\includegraphics[width=.9\textwidth]{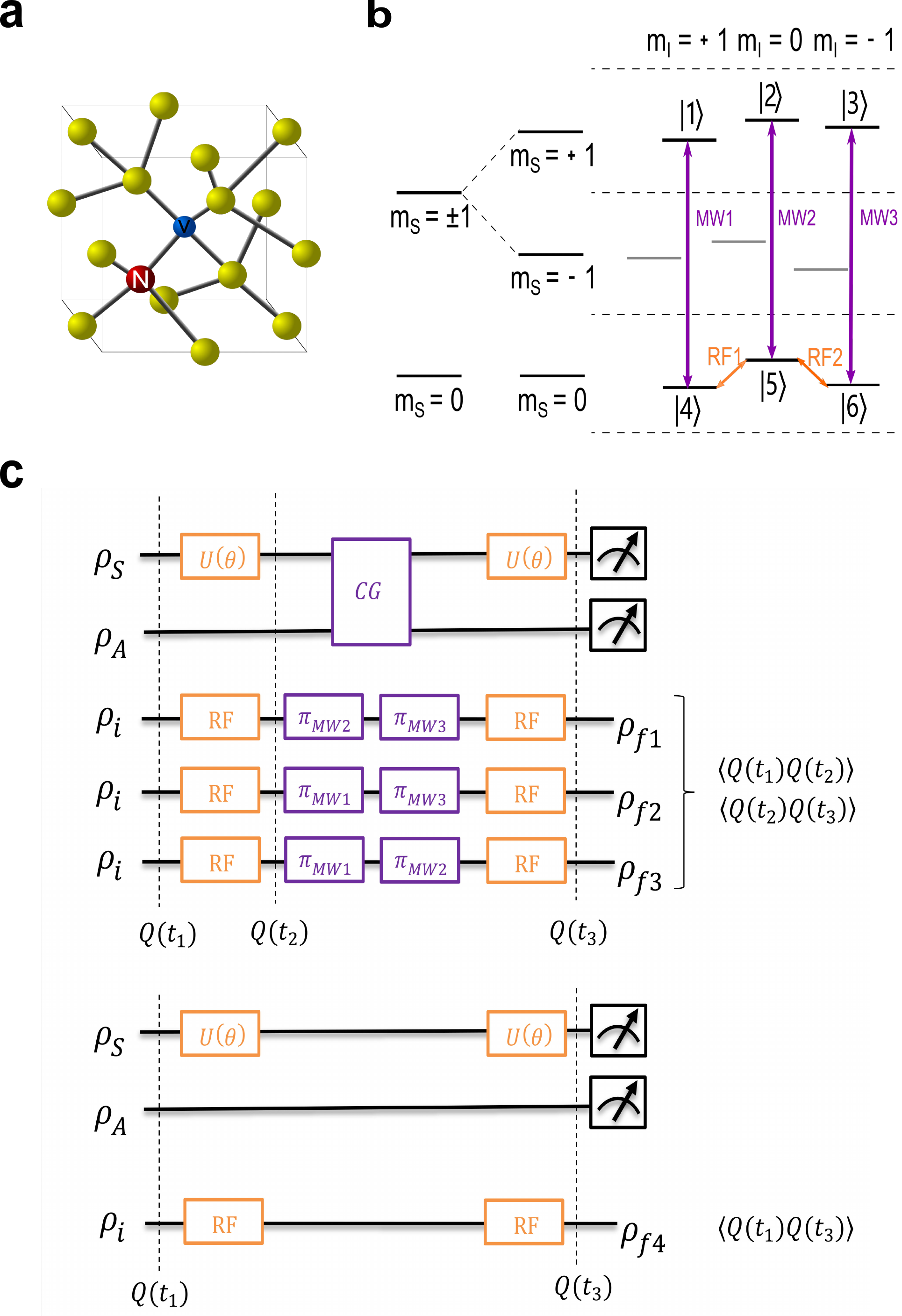}
\caption{(\textbf{a}) Atomic structure of NV center. (\textbf{b}) Energy levels of the NV center in the ground state. The $m_s$ and $m_I$ are the quantum states of electron spin and nuclear spin, respectively. MW1, MW2 and MW3 (blue arrows) are three selective microwave pulses to drive corresponding electron spin transitions. RF1 and RF2 (orange arrows) are two radio frequency pulses to drive transitions of the nuclear spin. (\textbf{c}) Experimental pulses sequence to realize INRM. The $\rho_{i}$ and $\rho_{fj}$, where $j=1,2,3,4$, are used to denote the initial and final state in the experiments. Controlled gate (CG) is realized through selective microwave pulses. Three-level Rabi rotation $U(\theta)$ on nuclear spin is realized by applying RF1 and RF2 simultaneously.}
\label{2}
\end{figure}

Our experiment was implemented on an NV center in diamond. As depicted in Fig.~\ref{2}(\textbf{a}), the NV center consists of a substitutional nitrogen atom with an adjacent vacancy site in the diamond crystal lattice. The three energy levels of ${}^{14}$N nuclear spin, $|1\rangle_{n}$, $|0\rangle_{n}$ and $|-1\rangle_{n}$, make up the three-level system. Two NV electron spin energy levels, $|1\rangle_{e}$ and $|0\rangle_{e}$, are chosen as an ancilla qubit. The Hamiltonian of the NV center is $H_{NV} = 2\pi(DS_{z}^{2}+\omega_{e}S_{z}+QI_{z}^{2}+\omega_{n}I_{z}+AI_{z}S_{z})$\cite{Wu_2018}, where $S_{z}$ and $I_{z}$ are corresponding spin operators of the electron and nuclear spin, $D = 2.87\,\rm{GHz}$ is the zero-field splitting of electron spin, $Q = -4.95\,\rm{MHz}$ is the nuclear quadrupolar interaction, and $A = -2.16\,\rm{MHz}$ is the hyperfine interaction. The experiment is performed on an optically detected magnetic resonance setup. A magnetic field of 512 G is applied along the NV symmetry axis ([111] crystal axis), yielding Zeeman frequencies $\omega_{e}$ and $\omega_{n}$. The total system utilized in our experiment is spanned by six energy levels, consisting of $|1\rangle_{e}\!|1\rangle_{n}$, $|1\rangle_{e}\!|0\rangle_{n}$, $|1\rangle_{e}\!|\textnormal{-}1\rangle_{n}$, $|0\rangle_{e}\!|1\rangle_{n}$, $|0\rangle_{e}\!|0\rangle_{n}$ and $|0\rangle_{e}\!|\textnormal{-}1\rangle_{n}$, relabeled by $|1\rangle$, $|2\rangle$, $|3\rangle$, $|4\rangle$, $|5\rangle$ and $|6\rangle$ as depicted in Fig.~\ref{2}(\textbf{b}).
 
The INRM is performed by the selective microwave pulses and postselecting the populations of the final states. The two microwave pulses, as shown in Fig.~\ref{2}(\textbf{c}), realize the controlled gate which flips the ancilla qubit when the system in two of the three states (see Appendix A for detailed discussion).
The three experiments correspond to the situation where the system is in three different states to ensure that the ancilla qubit is not flipped. The transition frequency of the microwave pulses is about 4.303 GHz, which is much different from the nuclear spin transition frequency on the megahertz order. These microwave pulses hardly flip the state of the nuclear spin. The non-invasiveness of the measurement is ensured by the postselecting the populations of the final states which indicate the case that the ancilla qubit is not flipped. The NV center is initialized to the state $|4\rangle$ via laser pulses\cite{Jacques_2009}. The postselection procedure is implemented by preserving the population of $|4\rangle$, $|5\rangle$ and $|6\rangle$ in the final states. Although the pulses in experiments will cause unwanted transitions due to finite off-resonance, these transitions can be removed through the postselection procedure. To verify the LG inequality, two time correlations of the measurements, $\langle Q(t_{i})Q(t_{j})\rangle$, need to be obtained.  Regarding the initialization as the first measurement, thus $Q(t_{1})=1$, the tested LG inequality is reduced to $\langle Q(t_{2})\rangle$ + $\langle Q(t_{2})Q(t_{3})\rangle$ -- $\langle Q(t_{3})\rangle\leq1$. The value of $\langle Q(t_{2})\rangle$ can be obtained from joint probability of measurement outcomes $P(q_2^{(l)},q_3^{(m)})$ when the controlled gates are performed. The value $\langle Q(t_{3})\rangle$ can be obtained from directly measurement without controlled gates intersected. The relationships between three terms in LG function and final state populations are shown below (see Appendix B for detailed calculations):
\begin{eqnarray}
\left\{
\begin{array}{lll}
\langle Q(t_{2})\rangle &= P_{4}^{1}+P_{5}^{1}+P_{6}^{1}+P_{4}^{2}+P_{5}^{2}+P_{6}^{2}\\
&-P_{4}^{3}-P_{5}^{3}-P_{6}^{3}\\
\langle Q(t_{2})Q(t_{3})\rangle &= P_{4}^{1}+P_{5}^{1}-P_{6}^{1}+P_{4}^{2}+P_{5}^{2}-P_{6}^{2}.\\
&-P_{4}^{3}-P_{5}^{3}+P_{6}^{3}\\
\langle Q(t_{3})\rangle &= P_{4}^{4}+P_{5}^{4}-P_{6}^{4}
\end{array}
\right.
\end{eqnarray}
where $P_{i}^{j}$ denotes the population of \textit{i}th energy level $\left.|i\right\rangle$ of the final state $\rho_{fj}$ in Fig.~\ref{2}(\textbf{c}), where $j=1,2,3,4$.
 
\begin{figure}
       \centering
       \includegraphics[width=.8\textwidth]{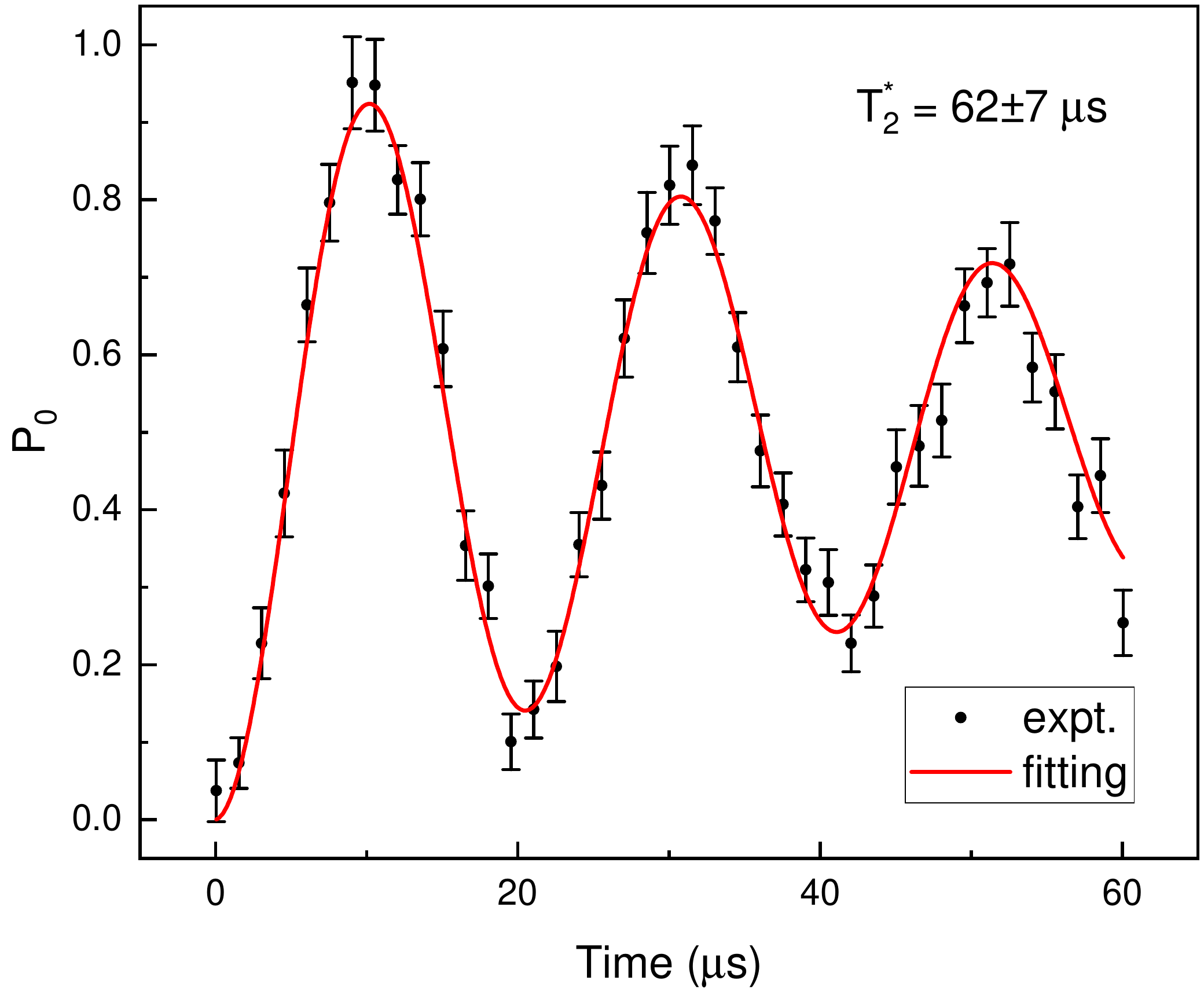}
       \caption{ Dephasing time of NV center. $P_0$ is the population of the electron spin state $|0\rangle_{e}$. The red line is fitted to the experiment data (black points with error bars). The decay time, $T_2^*$, of the free induction decay is measured to be 62(7) $\mathrm{\mu s}$. The error bars on the data points are the standard deviation from the mean.}
       \label{3}
\end{figure}
\begin{figure}
       \centering
       \includegraphics[width=1\textwidth]{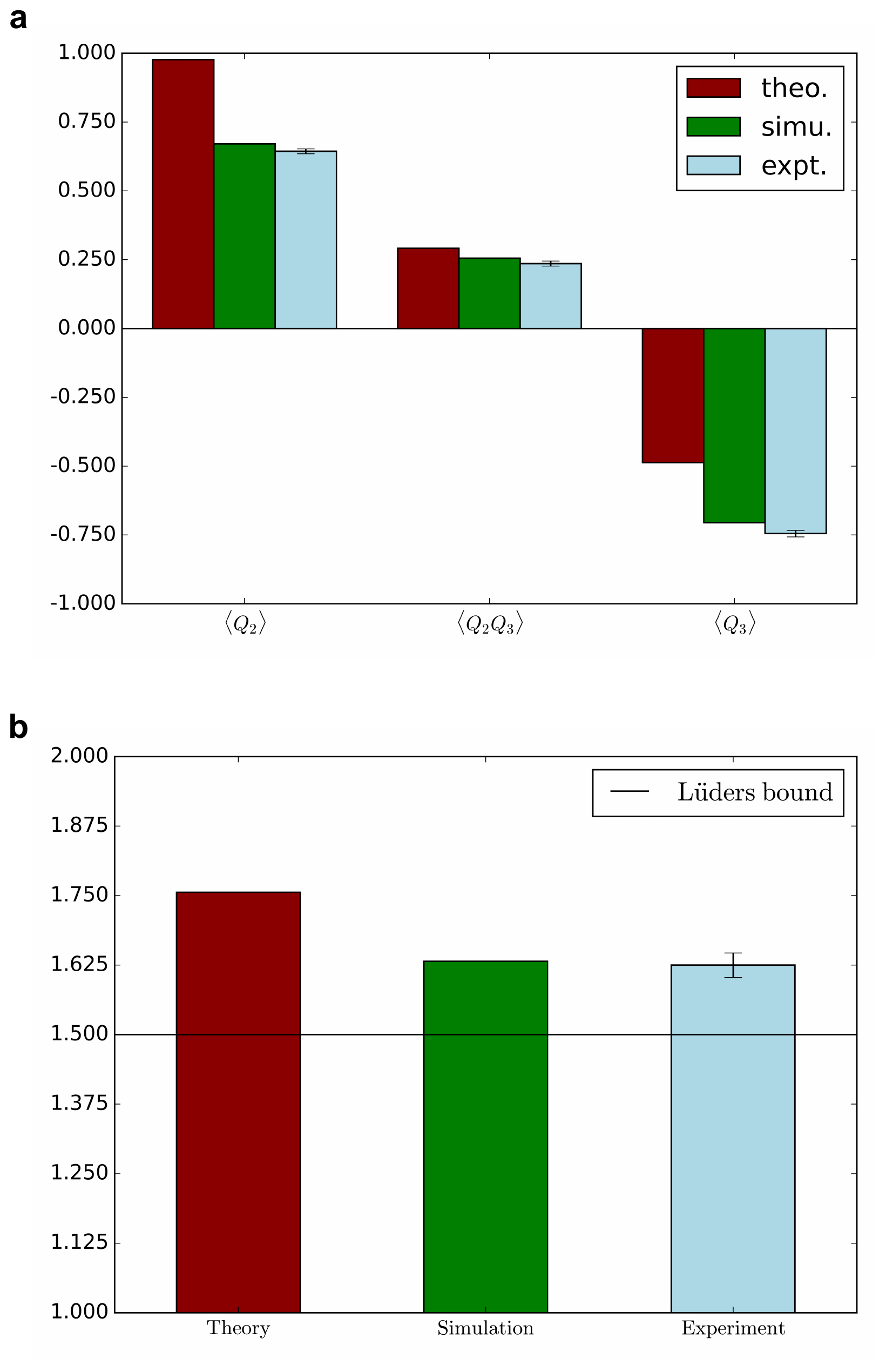}
       \caption{(\textbf{a}) Experimental results, simulation results and theoretical predictions of $\langle Q_{2}\rangle$, $\langle Q_{2}Q_{3}\rangle$ and $\langle Q_{3}\rangle$. (\textbf{b}) Experimental result (light blue bar), simulation result (green bar) and theoretical prediction (red bar) of the LG function $K_{3}$. The L$\mathrm{\ddot{u}}$ders bound (black line) is 1.5. The theoretical maximum violation in a three-level system is $K_{3}=1.756$ when $\theta=0.416\pi$. Considering dephasing noise and imperfect initialization of ancilla qubit, the corresponding simulation result is $K_{3}^{sim}=1.632$ and the experimental result is $K_{3}^{exp}=1.625\pm0.022$. The error bar of the experimental result is the standard deviation from the mean.}
       \label{4}
\end{figure}
 
The evolution of the rotation on the three-level system is $U(\theta) = e^{-i\theta S_{x}}$, where $\theta = \sqrt{2}\pi f_{rabi} t$ and $S_{x}$ is spin operator of spin-1 system.
The operation can be realized in rotating frame by applying two radio frequency pulses simultaneously on the nuclear spin.
The frequencies of the two pulses are $\omega_{45}$ , the resonance frequency between level $|4\rangle$ and $|5\rangle$, and $\omega_{56}$, the resonance frequency between level $|5\rangle$ and $|6\rangle$.
$f_{rabi}$ is the Rabi frequency which is proportional to strength of radio frequency pulses, which is set to $20\,\rm{kHz}$.
In theory\cite{Budroni_2014},
\begin{eqnarray}
\left\{
\begin{array}{lll}
\langle Q_{2}\rangle=\frac{1}{2}+\rm{cos}(\theta)-\frac{1}{4}\rm{cos}(2\theta)\\
\langle Q_{2}Q_{3}\rangle=\frac{1}{2}+\rm{cos}(2\theta)-\frac{1}{4}\rm{cos}(4\theta).\\
\langle Q_{3}\rangle=\frac{1}{16}+\rm{cos}(\theta)-\frac{1}{16}\rm{cos}(4\theta)
\end{array}
\right.
\label{5}
\end{eqnarray}
The maximal value of $K_{3}$ should be obtained when $\theta = 0.416\pi$, corresponding evolution time under the Hamiltonian is $14.71\,\mathrm{\mu s}$.
A major factor affecting the experimental results is the dephasing noise of the electron spin.
Therefore, we utilized an isotopically purified ([$^{12}$C]=99.999\%) sample to implement our experiments.
The dephasing time of the electron spin $T_2^*$ is measured to be 62(7) $\mathrm{\mu s}$ as shown in Fig.~\ref{3}.
Under this condition, the experimentally obtained values of $\langle Q_{2}\rangle$, $\langle Q_{2}Q_{3}\rangle$ and $\langle Q_{3}\rangle$ for $\theta = 0.416\pi$ are shown in Fig.~\ref{4}(\textbf{a}). The corresponding values of $K_{3}$ are shown in Fig.~\ref{4}(\textbf{b}), where the red (left) bar is the ideal prediction according to Eq.~(\ref{5}), the green (middle) bar is the simulation result with dephasing noise and initialization imperfection, and the light blue (right) bar is the experimental result. At the point of the maximum violation, the experimental result is $K_{3}^{exp}=1.625\pm0.022$. The error, $\sigma=0.022$, is the standard deviation from the mean. This result exceeds the L$\mathrm{\ddot{u}}$ders bound of 1.5 with a $5\sigma$ level of confidence. Considering dephasing noise and imperfect initialization of ancilla qubit, the corresponding simulation result is $K_{3}^{sim}=1.632$ and it is agrees with the experimental result (see Appendix C for detailed calculations).

In summary, we have experimentally demonstrated violation of LGI in a single spin three-level system beyond the L$\mathrm{\ddot{u}}$ders bound using the INRM scheme. The difference in experimental result from theoretical value is due to dephasing noise and imperfection of initialization of ancilla qubit.
To improve the experimental result, the effect of dephasing noise can be suppressed by using the sample with long coherence time and the quantum control which can resist noise. Some other quantities such as the quantum witness can be measured for the further investigation of macrorealist theories\cite{Schild_2015}. We hope that our results in three-level system can extend to large scale system, paving the way to looking for macroscopic superposition state in high-dimensional system.

This work was supported by the National Key R$\&$D Program of China (Grants No. 2018YFA0306600 and No. 2016YFB0501603), the NNSFC (No. 81788101), the Chinese
Academy of Sciences (Grants No. QYZDY-SSW-SLH004 and No. QYZDB-SSW-SLH005), and Anhui Initiative in Quantum Information Technologies (Grant No. AHY050000). X. R. thanks the Youth Innovation Promotion Association of Chinese Academy of Sciences for the
support. Ya Wang thanks the Fundamental Research Funds for the Central Universities for the support.

\section{Appendix A: Characterizing the flip probability of the electron spin when nuclear spin is in a flip state}
 
\begin{figure}
       \centering
       \includegraphics[width=0.8\textwidth]{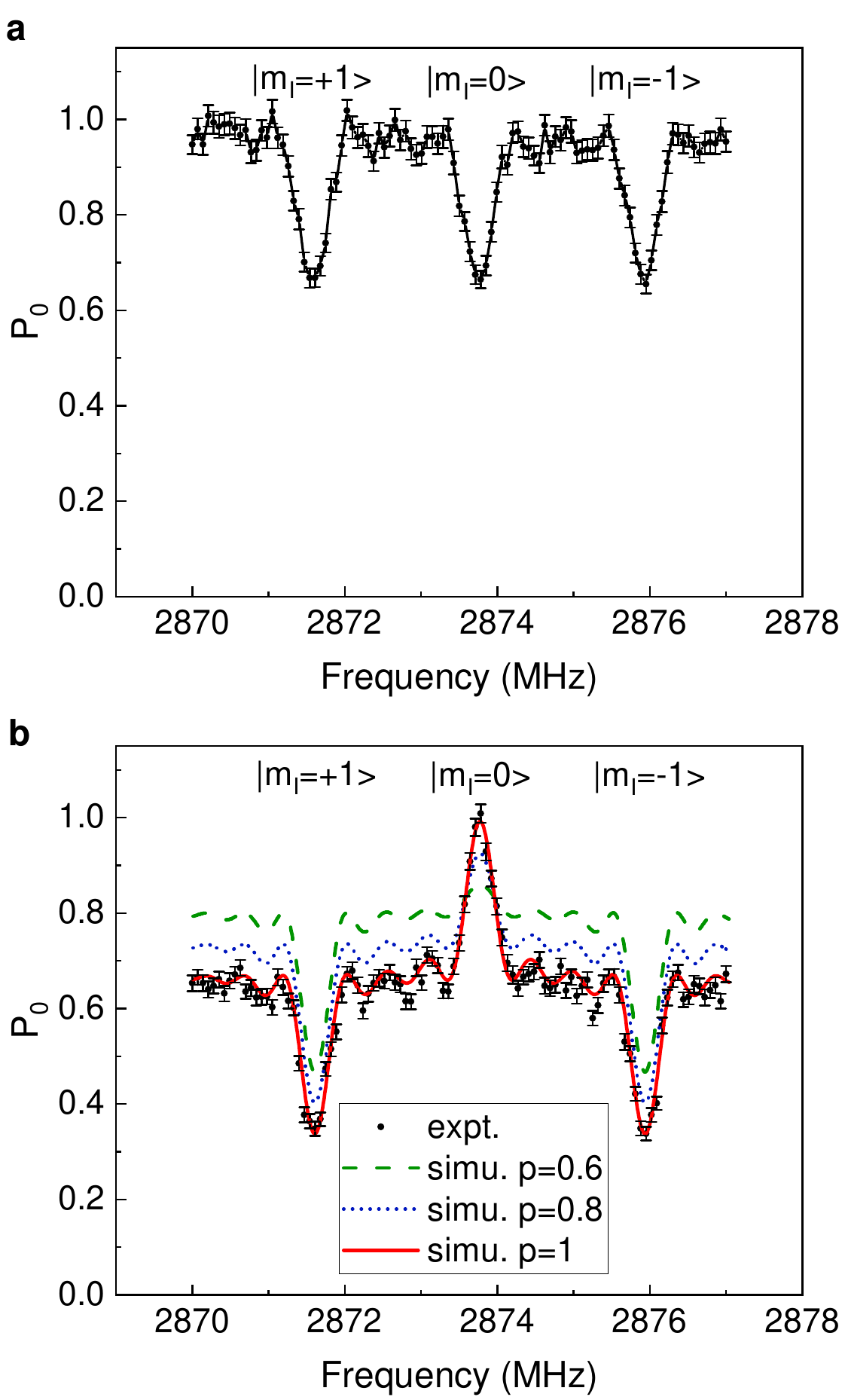}
       \caption{(\textbf{a}) The optically detected magnetic resonance (ODMR) spectrum of the electron spin before applying the controlled gate. Three peaks correspond to three different states of the nuclear spin. (\textbf{b}) The ODMR spectrum after applying the controlled gate. Black points with error bars is the experimental results. The red solid, blue dotted and green dashed lines correspond to the simulation results when the electron spin flip probabilities $p$ are 0.6, 0.8 and 1, respectively.}
       \label{fig5}
\end{figure}

We implemented experiments to show the probability $p$ of the ancilla qubit flipped when nuclear spin is in a flip state. Before applying the controlled gate, the electron spin is initialized to the $|m_S=0\rangle$ state. Thus the corresponding $|m_S=0\rangle$ state population $P_0$ is 1. When sweeping the frequency of the microwave pulse, we observed three peaks as shown in Fig. \ref{fig5}a corresponding to three different states of the nuclear spin. Fig. \ref{fig5}b shows the results after implementing the controlled gate.
 
\begin{figure}
       \centering
       \includegraphics[width=0.8\textwidth]{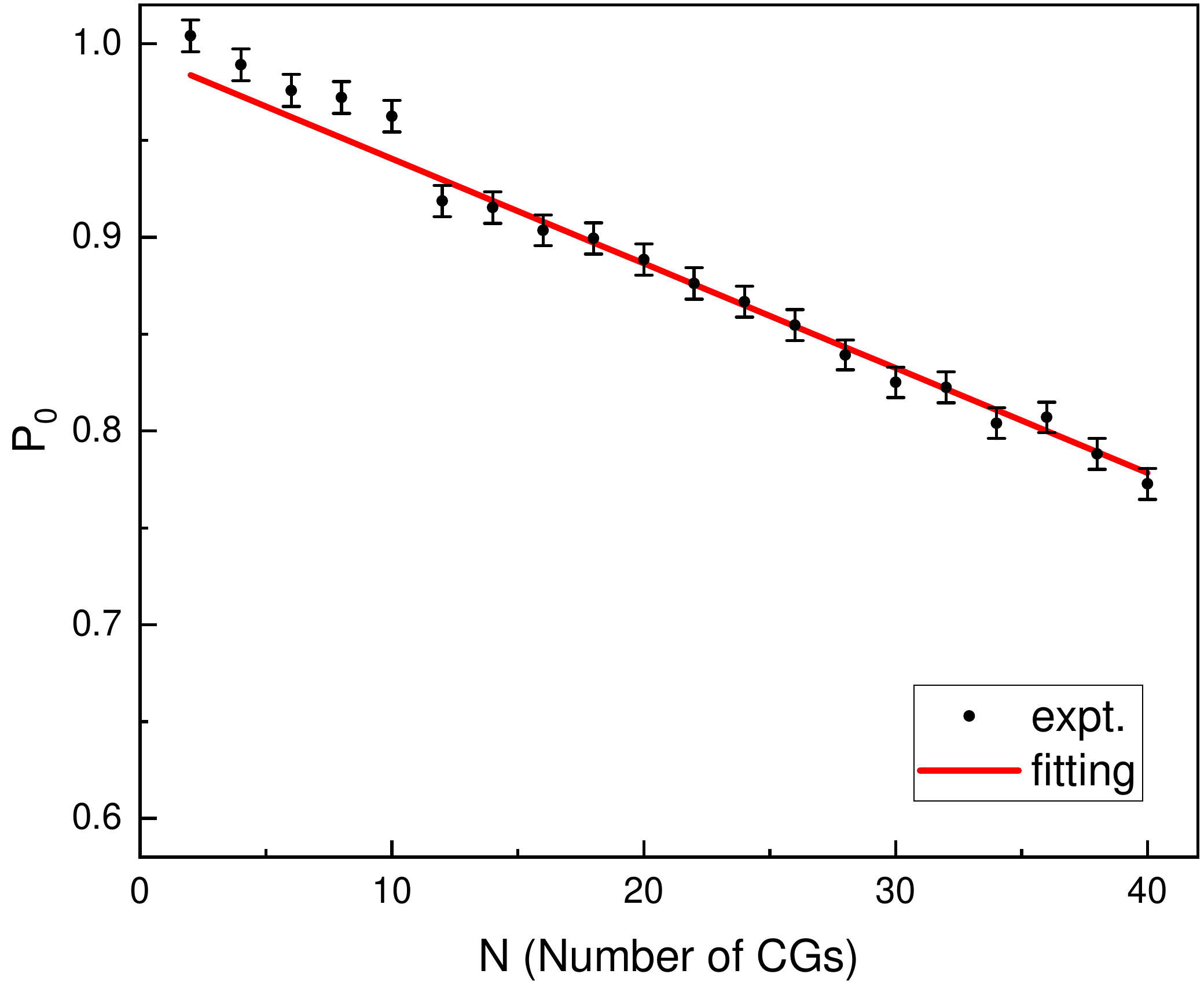}
       \caption{The $|m_S=0\rangle$ state population $P_0$ after applying controlled gates (CGs) repeatedly. The fitting (red line) agrees with the experimental data (black points with error bars). The error bars on the data points are the standard deviations from the mean.}
       \label{fig6}
\end{figure}

Compared to the simulation results, our results are quite different from the cases with probability 0.6 and 0.8, and agree well with the case of probability 1.
In order to obtain a more precise probability $p$, we performed an additional experiment which
applied the controlled gates repeatedly.
The dynamics of the quantum state was recorded.
The population of $|m_S=0\rangle$ state after repeated applications of controlled gates is shown in Fig. \ref{fig6}.
The probability p of $0.995\pm0.002$ can be obtained by linear fitting.

\section{Appendix B: LG function obtained from the populations of the final states}
\setcounter{equation}{0}
\setcounter{subsection}{0}
\renewcommand{\theequation}{B\arabic{equation}}
\renewcommand{\thesubsection}{B\arabic{subsection}}
At $t_2$, three controlled gates are applied corresponding to three nuclear spin states. We postselect the ancilla qubit to remain in the $|0\rangle_{e}$ state, which is not flipped. Therefore, the expectation value at $t_2$ is
\begin{equation}
\begin{aligned}
\langle Q(t_2)\rangle&=q_2^{(1)} (P_4^1+P_5^1+P_6^1 )+q_2^{(0)} (P_4^2+P_5^2+P_6^2 )\\
                     &+q_2^{(-1)} (P_4^3+P_5^3+P_6^3 ).
\end{aligned}
\end{equation}
The correlated expectation value is
\begin{equation}
\begin{aligned}
\langle Q(t_2)Q(t_3)\rangle&=q_2^{(1)} q_3^{(1)} P_4^1+q_2^{(1)} q_3^{(0)} P_5^1+q_2^{(1)} q_3^{(-1)} P_6^1\\
&+q_2^{(0)} q_3^{(1)} P_4^2+q_2^{(0)} q_3^{(0)} P_5^2+q_2^{(0)} q_3^{(-1)} P_6^2\\
&+q_2^{(-1)} q_3^{(1)} P_4^3+q_2^{(-1)} q_3^{(0)} P_5^3+q_2^{(-1)} q_3^{(-1)} P_6^3.
\end{aligned}
\end{equation}
In order to obtain $\langle Q(t_3)\rangle$, there is no need to apply the controlled gates at $t_2$. We still postselect the state where the ancilla qubit does not flip. Therefore, the expectation value at $t_3$ is
\begin{equation}
\langle Q(t_3)\rangle=q_3^{(1)} P_4^4+q_3^{(0)} P_5^4+q_3^{(-1)} P_6^4.
\end{equation}
 
\section{Appendix C: Imperfections Considered in the Simulation Results}
The imperfect polarization and the decoherence of the electron spin contribute to the difference between ideal and simulation results.
At the magnetic field strength of 512 G, the spin state of the NV center is effectively polarized to $|m_S=0, m_I=+1\rangle$.
The polarization of electron spin and nuclear spin are 95\% and 98\%.
Our experiment was implemented on a NV center in diamond which was isotopically purified ([$^{12}$C]=99.999\%).
The dephasing time of the electron spin, $T_2^*$ is measured to be 62(7) $\mathrm{\mu s}$.
The experiments suffered from the dephasing noise during pulse sequences.
There will be one more dephasing noise term, $H_{noise} = 2\pi\delta_0S_z$, in the practice Hamiltonian.
The dephasing noise, $\delta_0$, mainly comes from the Overhauser field (due to the interaction with the nuclear spin bath), the magnetic field fluctuation and the instability of the microwave frequency.
The $\delta_0$ satisfies a Gaussian distribution $P_0(\delta_0)=\exp(-\delta_0^2/2\sigma^2)/(\sigma\sqrt{2\pi})$ with $\sigma=1/(\sqrt{2}\pi T_2^*)$\cite{Rong_2015}.
The simulation results are obtained with the consideration of initial polarization and the dephasing noise distribution $P_0(\delta_0)$.

 
 
 
\end{document}